# Enhanced superconductivity in F-doped $LaO_{1-x}F_xBiSSe$

Xiao-Chuan Wang[1], Dong-Yun Chen[1], Qi Guo[1], Jia Yu[1], Bin-Bin Ruan[1], Qing-Ge Mu[1], Gen-Fu Chen[1,2], Zhi-An Ren[1,2]*

[1] Institute of Physics and Beijing National Laboratory for Condensed Matter Physics, Chinese Academy of Sciences, Beijing 100190, China

[2] Collaborative Innovation Center of Quantum Matter, Beijing 100190, China

* Email: renzhian@iphy.ac.cn

**Abstract**

The layered compound LaOBiSSe was successfully synthesized by a two-step solid-state reaction method, which is isostructural with the quaternary compound $LaOBiS_2$ indicated by the powder X-ray diffraction characterization, and the crystal lattice parameters are slightly expanded by the substitution of Se atoms. Electrical transport measurement shows that LaOBiSSe exhibits semiconducting behavior similar with the parent compound $LaOBiS_2$, while by F-doping the samples can be much more easily turned to be metallic and superconducting at low temperature with a small amount of F. Under the optimal carrier doping concentration of F ~ 0.5, the superconducting critical temperature of $LaO_{0.5}F_{0.5}BiSSe$ is clearly enhanced up to a value of 3.8 K than the 3.2 K of $LaO_{0.5}F_{0.5}BiS_2$.

## 1. Introduction

Chemical compounds with layered crystal structure are of great interest in the exploration for novel superconductors, and two types of high critical temperature (high-$T_c$) superconductivity have been found in copper oxides and iron pnictides that both possess layered lattices [1-2], in which the low dimensionality results in strong electronic correlation which is believed to enhance the unconventional high-$T_c$ superconductivity. Recently some $BiS_2$-based layered compounds were reported to be superconducting at low temperatures, such as $Bi_4O_4S_3$, $LnO_{1-x}F_xBiS_2$ (Ln = lanthanide elements) and $Sr_{1-x}La_xFBiS_2$ *etc.* [3-27]. The crystal structures of all these superconductors are generally composed of alternative stacking $BiS_2$-type layers and other insulating blocking layers, and the $BiS_2$-layers are believed to be responsible for the occurrence of superconductivity. For the typical $LaOBiS_2$ type quaternary compound which resembles a similar crystal structure with the iron-based LaOFeAs high-$T_c$ superconductors, superconductivity can be realized through electron-carrier doping by $F^-$ at the $O^{2-}$ atomic site, or $Ti^{4+}$ at the $La^{3+}$ site, and the superconducting transition temperature $T_c$ is usually around 3 K in samples synthesized from common solid state reaction method [4-5]. While interestingly, this $T_c$ can be dramatically increased up to around 10 K in samples synthesized by a high-pressure annealing method, or in samples measured under a high pressure around 1 GPa [6-10]. The enhancement of superconductivity mainly comes from the compression of the crystal structures in both cases, which indicates that structural modulation changes the superconducting $T_c$ significantly in this new $BiS_2$-based superconducting family.

Except for the physical pressure effect which always reduces the crystal lattices, chemical substitution with isovalent elements is another conventional method for structural modulation that is often used to modulate the superconducting $T_c$, and the crystal lattice can be either compressed or expanded by using isovalent elements with different atomic radius. A successful case is the Sm substitution for La in the $SmFeAsO_{1-x}F_x$ superconductor, which leads to a significant increase of $T_c$ from 26 K to 55 K [28]. In this study, the substitution with a larger atom of Se for S is used for structural modulation in the newly found $LaOBiS_2$ superconductor. The LaOBiSSe compound was successfully synthesized with a larger lattice, and superconductivity with a higher $T_c$ was achieved by F-doping.

## 2. Experimental

The F-doped polycrystalline LaOBiSSe and LaOBiS$_2$ samples were synthesized by a two-step solid state reaction method. At first, La$_2$S$_3$, La$_2$Se$_3$, Bi$_2$S$_3$ and Bi$_2$Se$_3$ were pre-sintered from La, Bi, S, and Se powders sealed in evacuated quartz tubes at 800 ºC for 12 hours, and these compounds were ground into powders for later use. Then the fine powders of La$_2$O$_3$, La$_2$S$_3$, La$_2$Se$_3$, Bi$_2$S$_3$, Bi$_2$Se$_3$, Bi, and LaF$_3$ were mixed together according to the nominal stoichiometric ratio of LaO$_{1-x}$F$_x$BiSSe (or LaO$_{1-x}$F$_x$BiS$_2$), ground thoroughly in an agate mortar and pressed into small pellets. All the preparation processes were carried out in an argon protected glove box. The pellets were sealed into evacuated quartz tubes and heated at 800 ºC for 50 hours followed by furnace cooling. The obtained samples were hard and black in color.

The samples were characterized by powder X-ray diffraction (XRD) method on a PAN-analytical x-ray diffractometer with Cu-Kα radiation from 20º to 80º at room temperature for crystal structure determination and phase analysis. The temperature dependence of resistivity was measured by the standard four-probe method using a PPMS system (Physical Property Measurement System, Quantum Design). The temperature dependence of DC magnetic susceptibility was measured during warming cycle by both zero-field cooling (ZFC) and field cooling (FC) method under an applied field of 10 Oe using an MPMS system (Magnetic Property Measurement System, Quantum Design).

## 3. Results and discussion

In Fig.1 the powder XRD patterns of all the prepared BiSSe-based samples of LaO$_{1-x}$F$_x$BiSSe with x = 0 - 0.7 were plotted, together with the patterns of BiS$_2$-based samples of LaOBiS$_2$ and LaO$_{0.5}$F$_{0.5}$BiS$_2$ for structural comparison. The main reflection peaks of all patterns can be well indexed with the reported CeOBiS$_2$ type tetragonal crystal structure with the space group P4/nmm. By F-doping at the oxygen site, minor impurity phases of Bi$_2$Se$_3$, Bi$_2$O$_3$ can be observed. And the ratio of these impurity phases increases with the F-doping content x, which indicates that fluorine cannot be fully doped at the O-site, and the real F-doping level is a little smaller than the nominal one. The existence of impurity phases is mainly due to the loss of fluorine during the high temperature reaction, which breaks the chemical stoichiometry, similar with the F-doping in iron oxypnictides. For the undoped LaOBiSSe phase, the refined lattice parameters are a = 4.118 Å and c = 14.133 Å, which are both much larger than those of LaOBiS$_2$, for which a = 4.062 Å and c = 13.858 Å. This lattice expansion comes from the larger Se atomic radius than the S atom. The separation of the (102) and (004) peaks is a typical feature for F-doped LaOBiS$_2$ samples, which also happens for the F-doped LaOBiSSe compound as shown in the right panel of Fig. 1. Clearly, the F-doping at the O-site causes the crystal lattice of LaOBiSSe to shrink significantly along the c-axis, which can be seen from the obvious right shift of the (004) peaks for the doped samples. The nearly unchanged positions of the (110) peaks indicate little variation of the tetragonal crystal structure along a-axis.

The temperature dependence of resistivity for all samples was measured from 2 K to 300 K to study the electron-doping induced superconductivity, and the data are shown in Fig. 2. For the two BiS$_2$-based samples, the resistivity curves show semiconducting-like behavior at normal state within the measured temperature range, and the F-doped LaO$_{0.5}$F$_{0.5}$BiS$_2$ (optimal-doping) shows an onset superconducting transition at $T_c \sim 3.2$ K, these results are similar as reported data [4-5]. For the BiSSe-based samples, the curve of undoped LaOBiSSe shows a similar semiconducting behavior, while all the curves of F-doped samples (from x = 0.1 ~ 0.7) show metallic behavior at normal state, and almost linear temperature dependence between 50 K ~ 300 K, with sudden superconducting transitions at lower temperatures, these characteristics are obviously different from the BiS$_2$-based samples, which indicates the replacement of S by Se atom has a clear change on the Fermi level electronic structure, and makes the samples more metallic and much easier to be superconducting. The onset of $T_c$ increases with F-doping content x, and exceeds the one of LaO$_{0.5}$F$_{0.5}$BiS$_2$ when x = 0.3, and reaches a maximum $T_c \sim 3.8$ K for the sample with the nominal composition of LaO$_{0.5}$F$_{0.5}$BiSSe, then slowly decreases with further F-doping. These results indicate that the LaOBiSSe is much easier to be turned metallic by electron doping, and superconductivity is enhanced in the LaOBiSSe compound by F-doping than in the LaOBiS$_2$ compound.

The temperature dependence of DC magnetic susceptibility for the superconducting samples of LaO$_{1-x}$F$_x$BiSSe (x = 0.3 - 0.7) was also measured in order to confirm the bulk nature of the superconducting behavior. Fig. 4 shows the susceptibility data between the temperature range of 2 K ~ 6 K measured with both field-cooling (FC) and zero-field-cooling (ZFC) methods at a magnetic field of 10 Oe. The diamagnetic transition clearly indicates the occurrence of superconductivity. The estimated superconducting shielding volume fraction from ZFC data is close to 33% for the sample with x = 0.4, which is similar with the F-doped $LaOBiS_2$ superconductors. The onset of the magnetic $T_c$ increases with increasing F content x, and reaches the highest of $T_c$ ~ 3.7 K for the samples of x = 0.4 and 0.5, then decreases with further increasing x. This doping behavior is consistent with the F-doped $LaOBiS_2$ superconductors but with an enhanced $T_c$.

To further elucidate the effect of F-doping and structure modulation in the LaOBiSSe compound, the lattice parameters 'a' and 'c' of all samples were refined from the XRD patterns and plotted in Fig. 4(a), and the onset $T_c$ determined from resistivity vs. the nominal F-content x was plotted in Fig. 4(b). It can be seen that with the increase of x, the crystal structure has a large shrinkage along the c-axis but little change along a-axis. When x > 0.4, the shrinking along c-axis becomes slower; this indicates that the doping becomes difficult when the F concentration increases, and the real doping level is smaller than the nominal x. This is consistent with that the ratio of impurity phases increases as x increases. The onset $T_c$ also rises quickly with the F-content x, while after x > 0.4, the change of $T_c$ becomes slower. The optimal doping level is at x = 0.5 with a $T_c$ ~ 3.8K, after that, the $T_c$ starts to decrease with more F-doping.

Here we note that the Se-substitution was independently studied in the layered $Bi_4O_4S_3$ superconductor which suppresses the superconductivity [29], and the fully Se-substituted $LaO_{0.5}F_{0.5}BiSe_2$ superconductor was also reported with a lower onset $T_c$ of 2.6 K [30]. Therefore the enhancement of superconductivity in this LaOBiSSe compound is quite interesting and worth to be further studied.

## 4. Conclusions

In summary, a layered compound LaOBiSSe was synthesized by solid-state reaction method. By F-doping at the O site which induces electron carriers, this compound can be much easily turned into superconducting with a higher $T_c$ than the F-doped LaOBiS$_2$ superconductors. The F-doping effect was systematically studied, which introduce electron carriers and causes lattice shrinkage mainly along the c-axis of the crystal lattice, and lead to a highest $T_c$ of 3.8 K in the optimally doped LaO$_{0.5}$F$_{0.5}$BiSSe superconductor.

**Acknowledgements**

The authors are grateful for the financial support from the National Basic Research Program of China (973 Program, No. 2010CB923000 and 2011CBA00100) and the Strategic Priority Research Program of the Chinese Academy of Sciences (No. XDB07000000).

**References:**

[1] Bednorz J G and Muller K A 1986 *Z. Phys. B-Condens. Mat.* 64 189
[2] Kamihara Y, Watanabe T, Hirano M and Hosono H 2008 *J. Am. Chem. Soc.* 130 3296
[3] Kotegawa H, Tomita Y, Tou H, Izawa H, Mizuguchi Y, Miura O, Demura S, Deguchi K and Takano Y 2012 *J. Phys. Soc. Jpn.* 81 103702
[4] Mizuguchi Y, Demura S, Deguchi K, Takano Y, Fujihisa H, Gotoh Y, Izawa H and Miura O 2012 *J. Phys. Soc. Jpn.* 81 114725
[5] Awana V P S, Kumar A, Jha R, Singh S K, Pal A, Shruti, Saha J and Patnaik S 2013 *Solid State Commun.* 157 21
[6] Selvan G K, Kanagaraj M, Muthu S E, Jha R, Awana V P S and Arumugam S 2013 *Phys. Stat. Sol. Rapid Res. Lett.* 7 510
[7] Wolowiec C T, White B D, Jeon I, Yazici D, Huang K and Maple M B 2013 *J. Phys. Condens. Matter* 25 422201
[8] Wolowiec C T, Yazici D, White B D, Huang K and Maple M B 2013 *Phys. Rev. B* 88 064503
[9] Deguchi K, Mizuguchi Y, Demura S, Hara H, Watanabe T, Denholme S J, Fujioka M, Okazaki H, Ozaki T, Takeya H, Yamaguchi T, Miura O and Takano Y 2013 *Europhys. Lett.* 101 17004
[10] Kajitani J, Deguchi K, Omachi A, Hiroi T, Takano Y, Takatsu H, Kadowaki H, Miura O and Mizuguchi Y 2014 *Solid State Commun.* 181 1
[11] Mizuguchi Y, Fujihisa H, Gotoh Y, Suzuki K, Usui H, Kuroki K, Demura S, Takano Y Izawa H and Miura O 2012 *Phys. Rev. B* 86 220510
[12] Singh S K, Kumar A, Gahtori B, Shruti, Sharma G, Patnaik S and Awana 2012 V. P. S. *J. Am. Chem. Soc.* 134 16504
[13] Takatsu H, Mizuguchi Y, Izawa H, Miura O and Kadowaki H 2012 *J. Phys. Soc. Jpn.* 125002
[14] Tan S G, Li L J, Liu Y, Tong P, Zhao B C, Lu W J and Sun Y P 2012 *Physica C* 483 94
[15] Demura S, Mizuguchi Y, Deguchi K, Okazaki H, Hara H, Watanabe T, Denholme S J, Fujioka M, Ozaki T, Fujihisa H, Gotoh Y, Miura O, Yamaguchi T, Takeya H and Takano Y 2013 *J. Phys. Soc. Jpn.* 82 033708
[16] Jha R, Kumar A, Singh S K and Awana V P S 2013 *J. Appl. Phys.* 113 056102
[17] Lee J, Stone M B, Huq A, Yildirim T, Ehlers G, Mizuguchi Y, Miura O, Takano Y, Deguchi K Demura S and Lee S H 2013 *Phys. Rev. B* 87 205134

[18] Lei H C, Wang K F, Abeykoon M, Bozin E S and Petrovic C 2013 *Inorg. Chem.* 52 10685
[19] Li B, Xing Z W and Huang G Q 2013 *Europhys. Lett.* 101 47002
[20] Lin X, Ni X X, Chen B, Xu X F, Yang X X, Dai J H, Li Y K, Yang X J, Luo, Y K, Tao Q, Cao G H and Xu Z A 2013 *Phys. Rev. B* 87 020504
[21] Nagao M, Demura S, Deguchi K, Miura A, Watauchi S, Takei T, Takano Y, Kumada N and Tanaka I 2013 *J Phys. Soc. Jpn.* 82 113701
[22] Wan X G, Ding H C, Savrasov S Y and Duan C G 2013 *Phys. Rev. B* 87 115124
[23] Yazici D, Huang K, White B D, Jeon I, Burnett V W, Friedman A J, Lum I K, Nallaiyan M, Spagna S and Maple M B 2013 *Phys. Rev. B* 87 174512
[24] Jha R and Awana V P S 2014 *J. Supercond. Novel Magn.* 27 1
[25] Jha R, Kishan H and Awana V P S 2014 *J. Appl. Phys.* 115 013902
[26] Nagao M, Miura A, Demura S, Deguchi K, Watauchi S, Takei T, Takano Y, Kumada N and Tanaka I 2014 *Solid State Commun.* 178 33
[27] Xing J, Li S, Ding X X, Yang H and Wen H H 2012 *Phys. Rev. B* 86 214518
[28] Ren Z A, Lu W, Yang J, Yi W, Shen X L, Li Z C, Che G C, Dong X L, Sun L L, Zhou F and Zhao Z X 2008 *Chin. Phys. Lett.* 25 2215
[29] Jha R and Awana V P S 2014 *Physica C* 498 45
[30] Maziopa A K, Guguchia Z, Pomjakushina E, Pomjakushin V, Khasanov R, Luetkens H, Biswas P, Amato A, Keller H and Conder K 2013 arXiv:1310.8131

**Figure captions:**

Fig. 1. X-ray powder diffraction patterns of the F-doped LaOBiSSe and LaOBiS$_2$ samples.

Fig. 2. Temperature dependence of normalized resistivity for the F-doped LaOBiSSe and LaOBiS$_2$ samples.

Fig. 3. Temperature dependence of DC susceptibility for the F-doped LaOBiSSe samples.

Fig. 4. The relationship between the lattice parameter a and c, the superconducting $T_c$ and the F-doping content x.

Fig.1

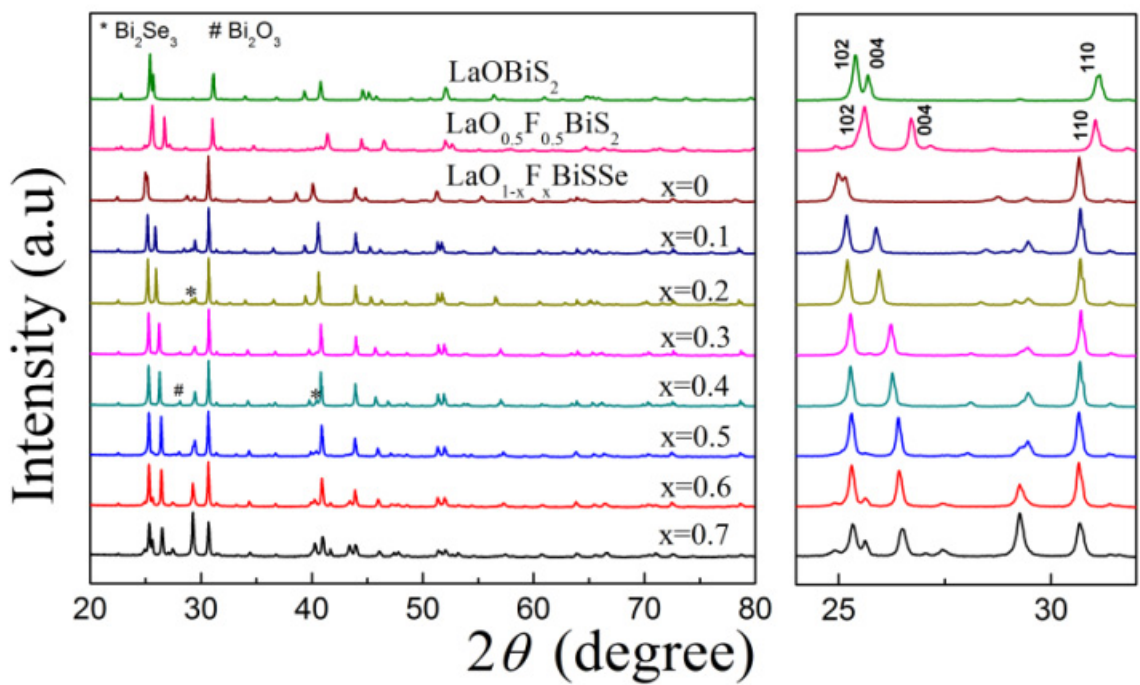

Fig.2

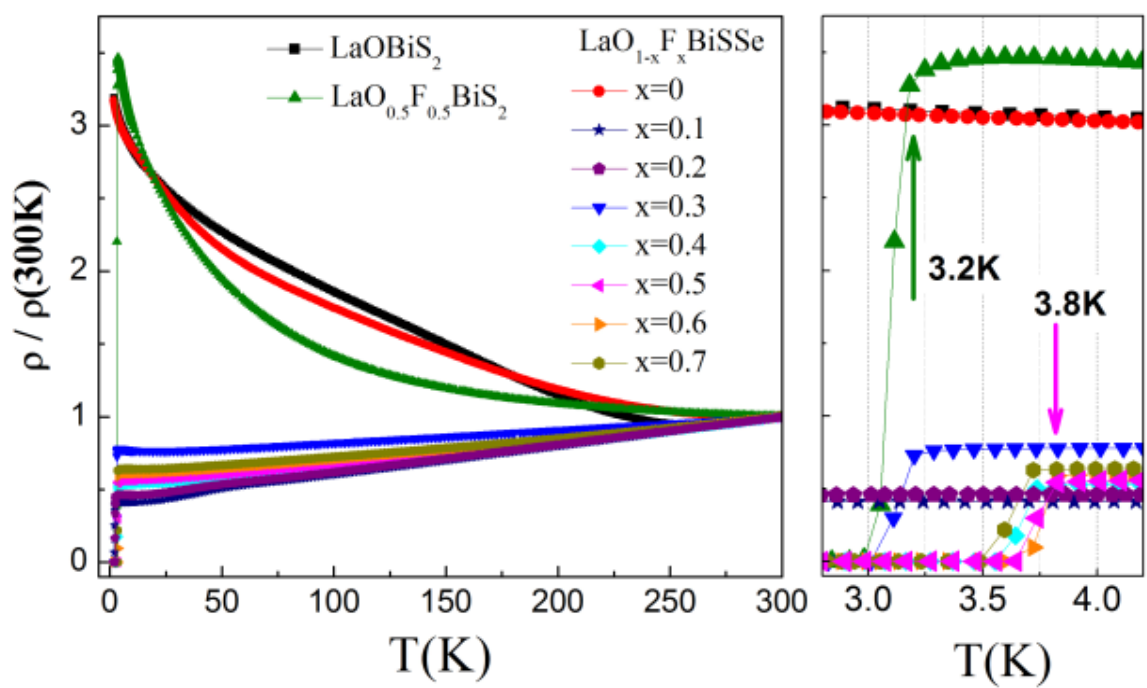

Fig.3

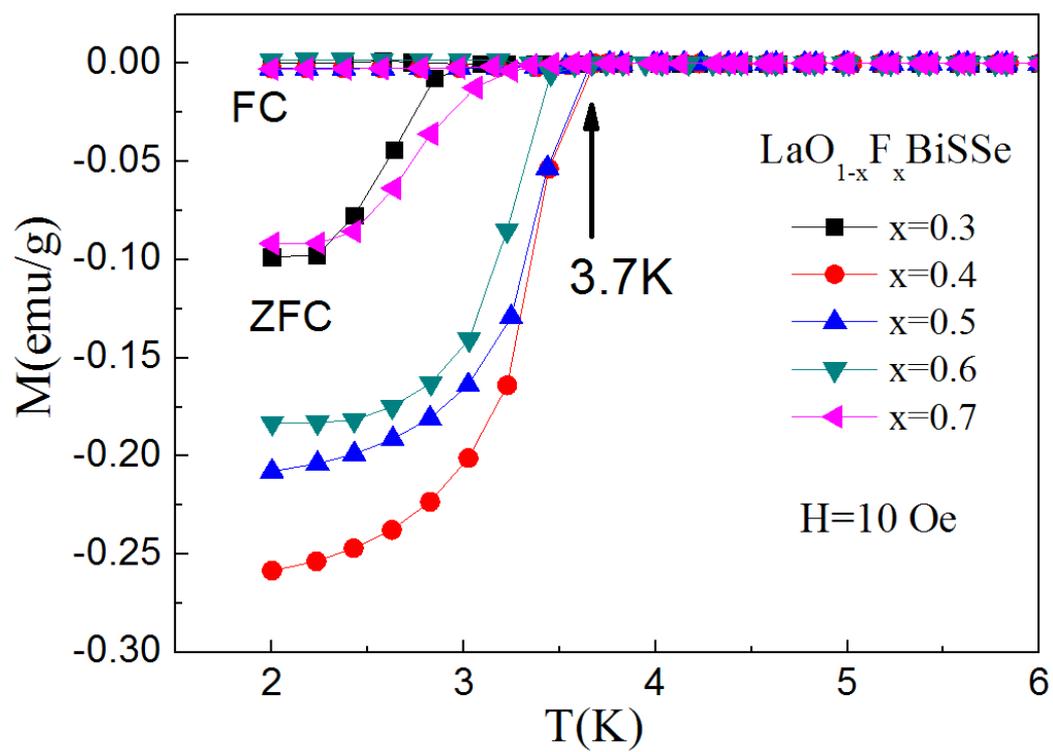

Fig.4

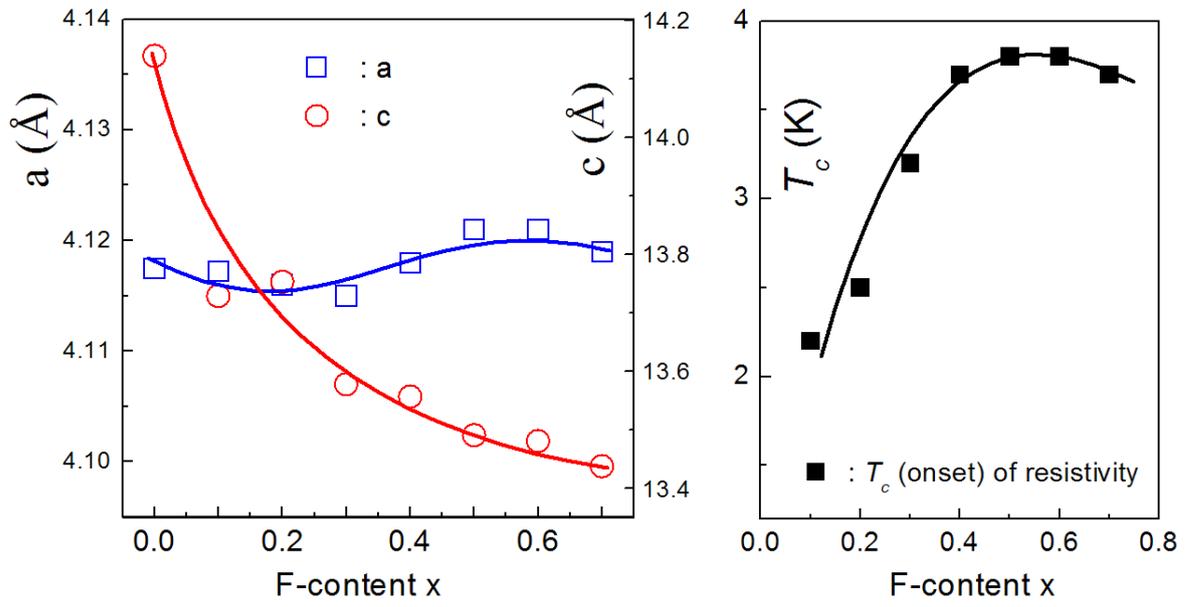